\documentclass{ws-acs}
\usepackage{cite}

\begin{document}

\markboth{D. Helbing et al.}{BioLogistics and the Struggle for
Efficiency: Concepts and Perspectives}

%
\catchline{}{}{}{}{}
%

\title{BioLogistics and the Struggle for Efficiency: Concepts and Perspectives}

\author{
Dirk Helbing$^{1,2,3}$, Andreas Deutsch$^{4}$, Stefan Diez$^{5}$,
Karsten Peters$^{4}$, Yannis Kalaidzidis$^{5}$, Kathrin
Padberg$^{4}$, Stefan L\"ammer$^{4}$, Anders Johansson$^{1}$, Georg
Breier$^{4}$, Frank Schulze$^{4}$, and Marino Zerial$^{5}$}

\address{
$^1$ ETH Zurich, Swiss Federal Institute of Technology,
Switzerland\\
$^2$ Santa Fe Institute, New Mexico, USA\\
$^3$ Collegium Budapest - Institute for Advanced Study, Hungary\\
$^4$ Dresden University of Technology, Germany\\
$^5$ Max Planck Institute of Molecular Cell Biology and Genetics,
Dresden, Germany}

\maketitle

\begin{history}
\received{(Day Month Year)}
\revised{(Day Month Year)}
\end{history}

\begin{abstract}
The growth of world population, limitation of resources, economic
problems and environmental issues force engineers to develop
increasingly efficient solutions for logistic systems. Pure
optimization for efficiency, however, has often led to technical
solutions that are vulnerable to variations in supply and demand,
and to perturbations. In contrast, nature already provides a large
variety of efficient, flexible and robust logistic solutions. Can we
utilize biological principles to design systems, which can flexibly
adapt to hardly predictable, fluctuating conditions? We propose a
bio-inspired ``BioLogistics'' approach to deduce dynamic
organization processes and principles of adaptive self-control from
biological systems, and to transfer them to man-made logistics
(including nanologistics), using principles of modularity,
self-assembly, self-organization, and decentralized coordination.
Conversely, logistic models can help revealing the logic of
biological processes at the systems level.
\end{abstract}


When the newly built Heathrow terminal 5 went into operation in
2008, it marked the beginning of a disaster: Thousands of lost
luggage items were piling up rapidly, passengers were delayed, etc.
It took about a week to fix the problem. In the complex logistic and
supply systems of today's highly connected, globalized world,
similar systemic failures occur again and again. Triggered by
insufficient responses to locally varying supplies or demands, a
problem can quickly spread over large parts of the system. Examples
for such problems range from blackouts of electric power grids up to
the current crises of the automotive industry and the financial
sector. This indicates that attempts to create highly efficient
systems are often compromised by the sensitivity of large man-made
structures to perturbations or varying demands, failures or attacks.

\par
Maximizing efficiency and profits often implies that redundancies
and safety margins are minimized under the constraint that certain,
yet acceptable failure rates are just kept. When connecting such
systems to form larger ones, coincidences of failures, and their
impact on a networked system are often underestimated. This is
particularly true for so-called ``complex systems'' (where
``complex'' is to be distinguished from ``complicated'' and from
``algorithmically complex'', i.e. computationally demanding). While
many large logistic systems have all three properties, the complex
systems we are focussing on are often characterized by non-linear
interactions of their elements and always by scalability issues.
That is, size matters for the resulting outcome and dynamic
behaviour of the system \cite{r1_anderson1}. Due to feedback loops
and reinforcement effects, natural fluctuations in the system may
trigger complex dynamics, instabilities, cascade failures and regime
shifts (i.e. transitions to a different state or operation mode).
For example, variations in the distances of vehicles can trigger the
breakdown of free flow, the formation of traffic jams (see video
SV1), and even {\em a drop in road capacity}, which may cause a
gridlock in large parts of the system \cite{r2_helbing2}.

\section{Learning from biology}

Generally, the task of logistics as a key element of our modern,
work-sharing economy is to make sure that required resources are
delivered in the right quality and quantity to the requesting
destination at the right time and acceptable costs. This involves
the organization, management, and engineering of suitable technical
systems. However, besides posing (NP-hard) problems, which may not
anymore be solvable in real-time by brute force supercomputing,
driving logistic systems towards maximum performance often drives
them to an instability threshold, causing undesirable breakdowns.
These are also known as ``slower-is-faster effects''
\cite{r2_helbing2} (see the previous example of traffic). It is
therefore evident that the struggle for efficiency in logistics is
closely tied to the above-mentioned complexity challenge and that
new solution strategies are needed.

\par
Biological transport systems and their organizational principles
could serve as an excellent source of inspiration for a variety of
new solutions. In fact, logistics in the sense of the organization,
coordination and optimization of material flows, is a ubiquitous
ingredient of biological systems, and bio-inspired approaches have
often solved engineering problems in the past: A considerable number
of natural structures and designs have been imitated under keywords
such as ``biomimicry'', ``biomimesis'', or ``bionics''
\cite{r3_ball3,r4_dickinson4,r5_wadhawan5,r6_kumar6} Moreover,
genetic algorithms and evolutionary optimization have been
successfully applied to many problems, where exact optimization was
not possible \cite{r7_passino7}. In contrast to previous
bio-inspired approaches, which were primarily focused on imitating
structural designs, we propose to concentrate on functional
principles now, particularly on issues of dynamics and adaptive
organization.

\section{The approach of BioLogistics}

Analogies between biological and man-made logistic systems on the
structural, functional, and dynamic level suggest a new,
multi-disciplinary research field of {\em ``Bio\-Logistics''}, which
connects biology with complexity science and engineering. The
BioLogistics approach includes the derivation of innovative logistic
solutions from successful strategies of nature as well as the
analysis of biological systems from the logistics point of view. Of
course, there are marked differences between biological systems and
man-made logistics in terms of items and quantities, building blocks
and mechanisms, designs and structures, topologies and scales,
dynamics and growth, or organization and control (see supplementary
table 1). Nevertheless, it turns out to be fruitful to exploit the
same vocabulary of design principles and properties such as
resource-efficiency, robustness, adaptability, and noise control.

\par
Biological logistic systems do an excellent job in using,
distributing, and recycling sparse resources. Moreover, their
efficiency and robustness to environmental perturbations are key to
their survival. Millions of years of evolution have created logistic
systems of such large variety and astonishing performance that one
can hope to reveal a multitude of yet undiscovered functional
designs and heuristic solutions. For example, cells have to produce
or import, correctly transport, localize and monitor the activity of
thousands of different molecules. The human body even represents a
``logistic universe'' comparable with the phenomenal complexity of
man-made global logistics: It manages to transport millions of
different materials (nucleic acids, proteins, lipids, carbohydrates
and metabolites) to different destinations, establishing billions of
molecular and cellular interactions (e.g. neuronal connections).
This is done with an incredibly low energy consumption of a 60-100
Watts bulb ($<$2,000 kCal/day) and at very high reliability.

\par
Understanding logistic processes in biology is, therefore, a new
challenge. The current progress in cell biology can make big
contributions to identifying molecular structures and regulation
mechanisms. While the components of biological processes are being
catalogued, we now need to unravel their dynamic interplay on the
systems level. For example, how do biological systems respond to or
even use variations in the concentrations of metabolites? A
logistics view could largely help to reveal these dynamic
interactions, and the systemic functions resulting from them. In
recognition of the similarities between biological and man-made
logistic systems, we will now introduce the BioLogistics approach by
highlighting some key aspects concerning the infrastructure and
organizational principles.

\section{Structure, function, and dynamic organization}

From a {\em structural} perspective, it is interesting to compare
technological networks realizing logistics (such as distribution
systems, vehicle or data traffic) with biological networks:
Remarkably, man-made and biological logistic networks share the same
notions of transport infrastructure (roads, railways or conveyor
belts on the one hand, a cytoskeleton at the cellular scale or
vascular system at the organism scale on the other). In both systems
we find mobile transport units and buffers for intermediate storage.
Moreover, design principles of volume- or area-covering transport
networks, respectively, seem to have universal features
\cite{r8_west8,r9_banavar9}. This is the case, because both
engineered and biological transport infrastructures minimize the use
of resources under functional and physical constraints. Both aim to
guarantee a reliable supply in all parts of the system, and to
maximize transport efficiency.

From a {\em functional} perspective, we find again astonishing
similarities between the two systems. For example, cellular traffic
along cytoskeletal tracks (microtubules or actin filaments) may be
compared with traffic along roads and highways. This shall be illustrated by 
Example 1.

{\small\sffamily \begin{itemize}
\item[] \textbf{Example 1: Intracellular traffic and crises in logistic transport}
\par
Most (eukaryotic) cells in nature comprise a cytoskeleton made up of
filaments (such as microtubules, which have a diameter of about 25
nanometre and a length of several tens of micrometres). Apart from
providing structural stability to the cell, these filaments serve as
(multi-lane) tracks for motor proteins, which facilitate the
long-range transport of various cargo (see video SV2). However, just
as in urban and highway traffic (see Fig.~\ref{fig1}a), high demand
may lead to traffic jams, suboptimal throughput and even collapse.
\par\begin{figure}[htbp]
\centerline{\psfig{file=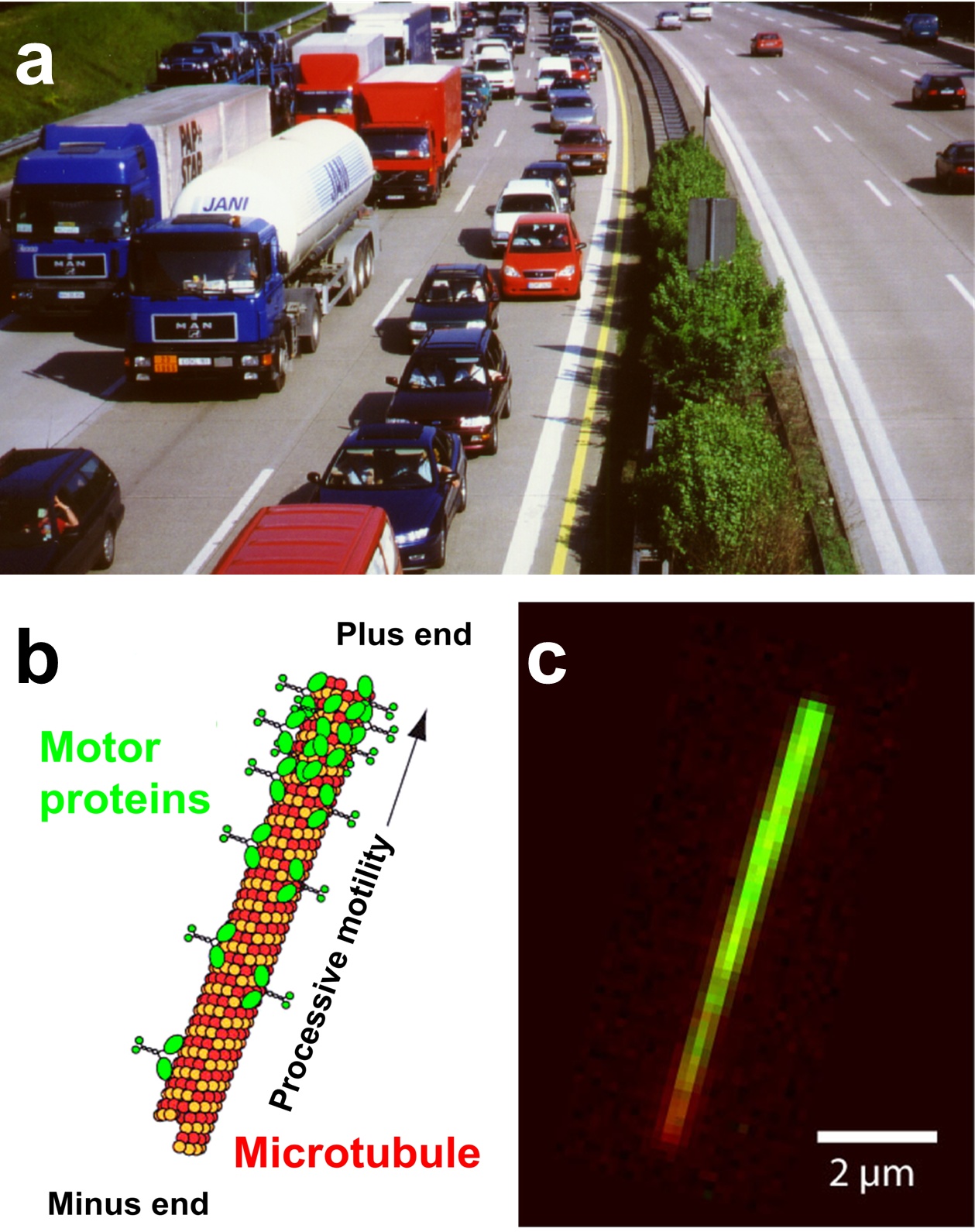,width=3in}}
\caption{(a) Traffic jams on a road (see also video SV1) . (b) Traffic jams along a microtubule
in schematic representation. Motor proteins
moving towards the microtubule plus end may form a concentration
gradient and thereby cause crowding effects. (c) Experimental
results for traffic jams along microtubules, obtained by fluorescence microscopy of red-labelled
microtubules and green-labelled motor proteins (adapted from
Ref.~\cite{r29_varga29}). See also video SV1. \label{fig1} }
\end{figure}
To avoid such states of crisis, cells have a number of control
mechanisms at their disposal: the speed of cargo carried by
molecular motors is influenced (i) by the composition of the
surrounding solution, i.e. dependent on how much ``fuel'' (usually
adenosine triphosphate, ATP) is provided, or whether there are
specific inhibitors present, (ii) by the actual molecular structure
of the motor itself, and (iii) by the ``road conditions''. As such,
optimal transport may be impeded by obstacles on the filament
surface \cite{r28_korten28} or under conditions of motor crowding
\cite{r29_varga29} (see Figs.~\ref{fig1}b+c). Microtubule associated
proteins (MAPs) can bind the microtubule surface independent of the
motors. Once bound, MAPs may block the motor binding sites and
influence the transport efficiency (i.e. cargo throughput). This
kind of influence is part of the natural machinery of transport
regulation \cite{r30_dixit30}. However, if the MAP-concentration
exceeds a certain threshold, transport may break down, leading the
system into a state of crisis. Understanding the mechanism behind
such logistic crises may also allow the development of new
therapeutic strategies for a number of diseases \cite{r31_aridor31}.

BioLogistics is particularly interested in analyzing the mechanisms,
whereby cells cope with and avoid congestion. Such biologically
inspired solutions will be useful for the improvement of man-made
systems. However, BioLogistics shall also work the other way around:
Logistic models can shed new light on biological processes and
reveal their underlying mechanisms and dynamics. For example, they
can help to understand the functional role that the crowding of
motor proteins towards the microtubule plus ends may play (see
Figs.~\ref{fig1}b+c).
\end{itemize}
}

Both technical and biological logistic
systems use multi-modal transport, i.e. a combination of different
transport modes such as truck and railway traffic in distribution
logistics, or diffusion and directed motion in the cell
\cite{r10_helenius10}. They perform the same functions like sorting,
storage and transport, which require mechanisms for item
identification, destination search, routing, carrier choice, and
delivery. Strategies for communication, coordination, and prioritization are
essential as well. Furthermore, biological and man-made logistics
seem to share pretty much the same goals, including cost efficiency,
high throughput, spatial coverage, and robustness. The most
important insights from biology, however, are expected from the
perspective of {\em dynamic organization}, and its smooth interplay
with the structural and functional level. As information technology
becomes more powerful, efficiency challenges in logistics are
increasingly addressed by a central control of production plans,
quantities, time schedules, delivery times, and routes. This,
however, may imply a low degree of flexibility with respect to
varying conditions, a considerable sensitivity to perturbations,
large administrative overheads and, sometimes, system breakdowns.
Biological systems, in contrast, do not perform optimization in a
strict sense. To understand why, we need to extend the notion of
efficiency to the successful survival of a complex system in an
uncertain and challenging environment. This includes the ability to
respond to changing demands or sudden changes, and the maintenance
of functionality under stressful conditions.

\section{Success strategies of nature}

Design principles of nature have to allow for flexible adaptation to
requirements such as growth, shrinkage, or reorganization. The
vascular system is a particularly well-known example, which has the
ability to adapt to varying metabolic needs and body growth (see 
Example 2). The underlying success strategies include
extensive recycling, self-organization, self-assembly
\cite{r11_zhang11,r12_li12} and self-repair. These properties are
largely based on local interactions, i.e. on decentralized control
approaches \cite{r13_kaneko13,r14_mikhailov14}. 

{\small\sffamily\begin{itemize}
\item[] \textbf{Example 2: Sorting and distribution in engineered and biological systems}
\par\begin{figure}[bp]
\centerline{\psfig{file=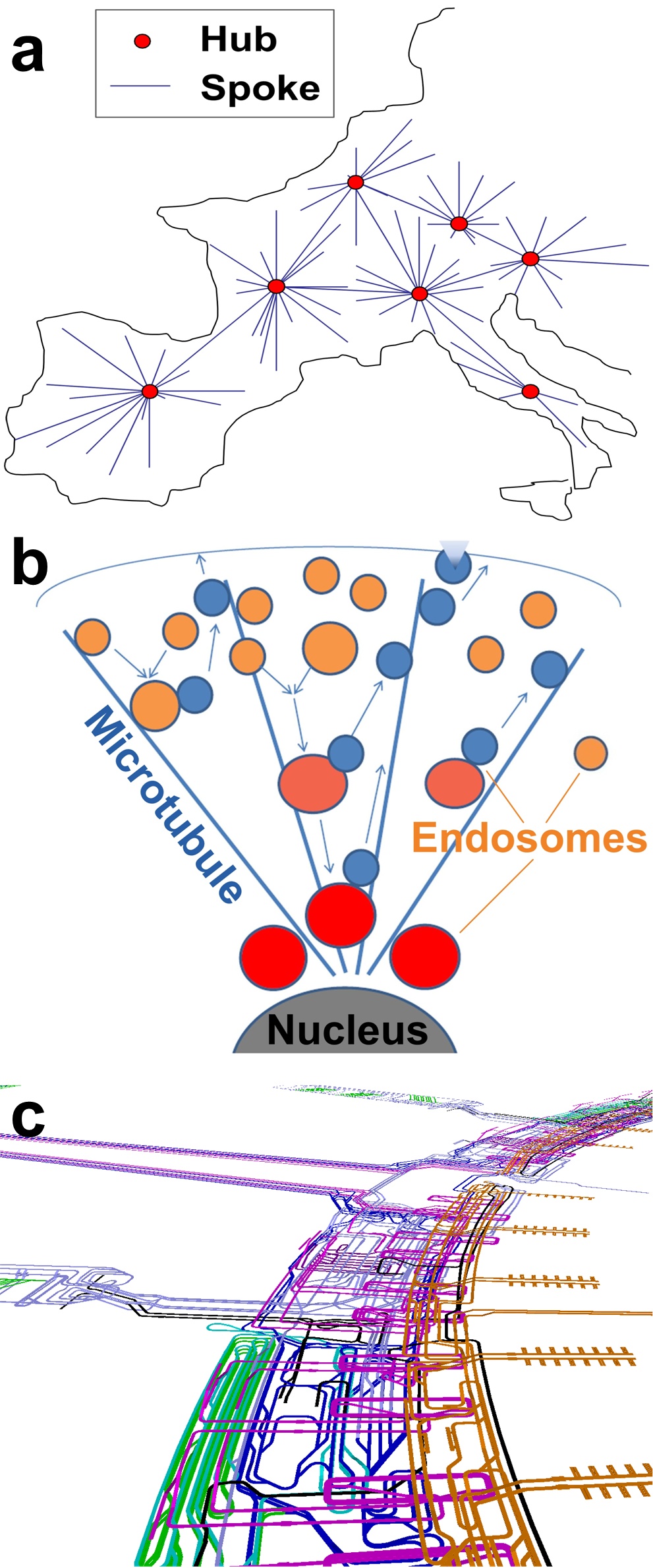,width=1.8in}} \caption{(a)
Illustration of a typical hub-and-spoke distribution network. (b)
Schematic picture of the main processes during endocytosis (see also video SV3). (c)
Illustration of a baggage handling system (BHS) used in airport
baggage logistics. \label{fig2}}
\end{figure}
In engineered distribution networks, transport and sorting are
usually well separated: A fundamental principle of logistic systems
is the use of economies of scale. As sorting is costly, sorting
costs per item are minimized by having large-scale sorters in hubs
(see Fig.~\ref{fig2}a). All cargo from a certain region is
transported to the hub, where it is sorted according to
destinations, products, qualities etc. Afterwards, it is shipped to
the destination, often in a containerized way, without performing
any other activity during the transport. 
\par
Biological systems, in
contrast, tend to perform several functions in parallel. Examples of
such multiplexing activities at the intra-cellular and
multi-cellular scale are provided by endocytic and vascular
networks. In a process called endocytosis, cells take up nutrients
and signalling molecules from the outside. Whereas several types of
cargo are transported to specialized compartments (lysosomes) for
digestion, others (e.g. some cargo receptors) together with their
containers can be recycled back to the surface of the cell to be
re-utilized. Recycling is thus an important feature of cellular
logistics. Cargo transport between the periphery and the cell centre
occurs concomitantly with cargo sorting and information exchange
(signal transmission to the nucleus). This parallelism allows cells
to carry out cargo transport and sorting on enormously compressed
temporal and spatial scales, and to avoid overcrowding in transport
processes towards the nucleus. The high logistic performance in the
cell is reached by a combination of several processes (see
Fig.~\ref{fig2}b and video SV3): 1. fusion of small endosomes
(``containers'') to form compact, larger ones, where cargo
accumulates; 2. fission of endosomes to partially separate cargo to
be transported towards the nucleus (for degradation) from cargo to
be moved towards the cell surface (for recycling); 3. non-uniform
motion, where periods of restricted movement alternate with periods
of long-range transport \cite{r32_rink32}. Iterative repetitions of
steps 1 and 2 during endosome movement facilitates to achieve high
sorting quality and to keep control of cargo destination. Besides
avoiding logistic problems like crowdedness and traffic jams, the
regulation of motility allows to trade off sorting quality and cargo
delivery time. It will be interesting to determine under which
conditions similar, bio-inspired solutions can outperform today's
logistic systems. Furthermore, the optimal degree of centrality vs.
de-centrality will be an important question to address. 

Vascular networks also fulfil several functions simultaneously such as
providing oxygen and nutrients to all cells in the body. They remove
waste products, support immune defence, and serve communication,
e.g. via growth factors and hormones, which are taken up by cells
through endocytosis. Moreover, they are highly adaptive to changing
transport demands, and they have many things in common with baggage
handling systems (see Fig.~\ref{fig2}c). Therefore, it seems natural
to transfer their design and operation principles to better adapt
baggage handling systems to varying transport volumes, changing
origin-destination relations, and required system extensions, when
the number of passengers increases.
\end{itemize}
}

Decentralized control approaches tend
to scale better in terms of required control resources than
centralized systems (see Example 3). Moreover, they are less
sensitive to failures of systems elements. Despite the prevalence of
local interaction mechanisms in biological systems, however,
decentralization is not a {\it general} rule, as the nervous system of
higher organisms shows. Apparently, shortcuts between remote parts
of a large, complex system are needed, when a local spreading of
coordinated behaviour would imply too large delays, which could
cause unstable dynamics, loss of control, or system collapse. 

To guarantee robustness, essential functions in biological systems are
implemented in redundant ways (i.e. there is often a
``plan B''). The body, for example, has alternative mechanisms of
transport or of energy storage and provision. While these are often
used in parallel to increase efficiency, in crises they may
substitute each other. This redundancy also supports a large degree
of autonomy and independence from specific environments and
resources. 

Last but not least, biological systems can cope well with
fluctuations: While logistic operation in man-made systems typically
fights fluctuations as undesired deviations from some planned
behaviour, biological systems often exploit fluctuations, e.g. for
short-range transport via diffusion, or for innovation and
evolutionary progress via mutations. Many functional states of
biological systems are reached by self-organization, which makes
them stable with respect to small fluctuations. In contrast, large
fluctuations (in demand, for example) tend to trigger transitions to
different system behaviours \cite{r14_mikhailov14} (e.g. from
metabolizing carbohydrates to burning fat). Many biological systems
use such transitions for the adaptation to changed conditions like
lower supply, and fluctuations are part of their functional design.
In particular, fluctuations can support a reorganisation of the
system \cite{r15_ueda15}.

\section{The challenge of putting BioLogistics to practice}

Could the above principles be exploited for a bio-inspired
logistics? We definitely think so. But how can we optimize
self-organizing systems that are prone to capacity breakdowns when
maximizing throughput? Example 3 describes an
application to urban traffic flow control, which uses randomly
occurring gaps in vehicle streams to serve other flow directions or
other modes of transport. As a result, vehicles can enter the
intersection according to a first-come-first-serve principle, when
traffic volumes are low. At higher traffic flows, however, traffic
lights form and serve {\em platoons} of vehicles, which is more
efficient \cite{r16_laemmer16}.

{\small\sffamily\begin{itemize}
\item[]\textbf{Example 3: Centralized optimization vs. self-control}
\par
When logistic or transport systems are designed and operated, it is
quite common today to solve a related, multi-criterial optimization
problem with powerful methods from Operations Research, considering
costs, delays, or other criteria. This involves the maximization of
a goal function. However, there are many possible ways of specifying
the goal (e.g. to minimize costs or to maximize throughputs), many
possible boundary conditions (reflecting the assumed variability of
prices and demands, or the distribution of transport volumes and
destinations), and different scenarios (regarding situations that
the system may face, e.g. certain failures). A system is usually
optimized for a specific goal and expected typical conditions, but
these conditions may never occur exactly. The goal may change as
well, which may imply a significantly reduced performance as
compared to the optimal operation that the system was designed for.
Moreover, the behaviour of complex systems is often hard to predict,
and it may react in a sensitive way to external control. Under such
conditions, a centralized feedback control (top-down approach),
often requiring a supercomputer, may not be the best approach: It
needs costly control structures and is sensitive to failures of the
central unit due to errors or attacks. For so-called
non-polynomially hard (NP-hard, computationally demanding) problems,
central performance is often insufficient for real-time optimization
of large systems, which may cause information overload, negligence
of potentially important information, and delay-induced
instabilities. A common way to overcome this problem is to restrict
the solution space, i.e. to introduce standard solutions suited for
typical situations. This, however, reduces variability, which in
turn affects the flexibility and adaptability to local or changing
conditions. As a consequence, this may harm the efficiency and
robustness of a system. 
\par\begin{figure}[htbp]
\centerline{\psfig{file=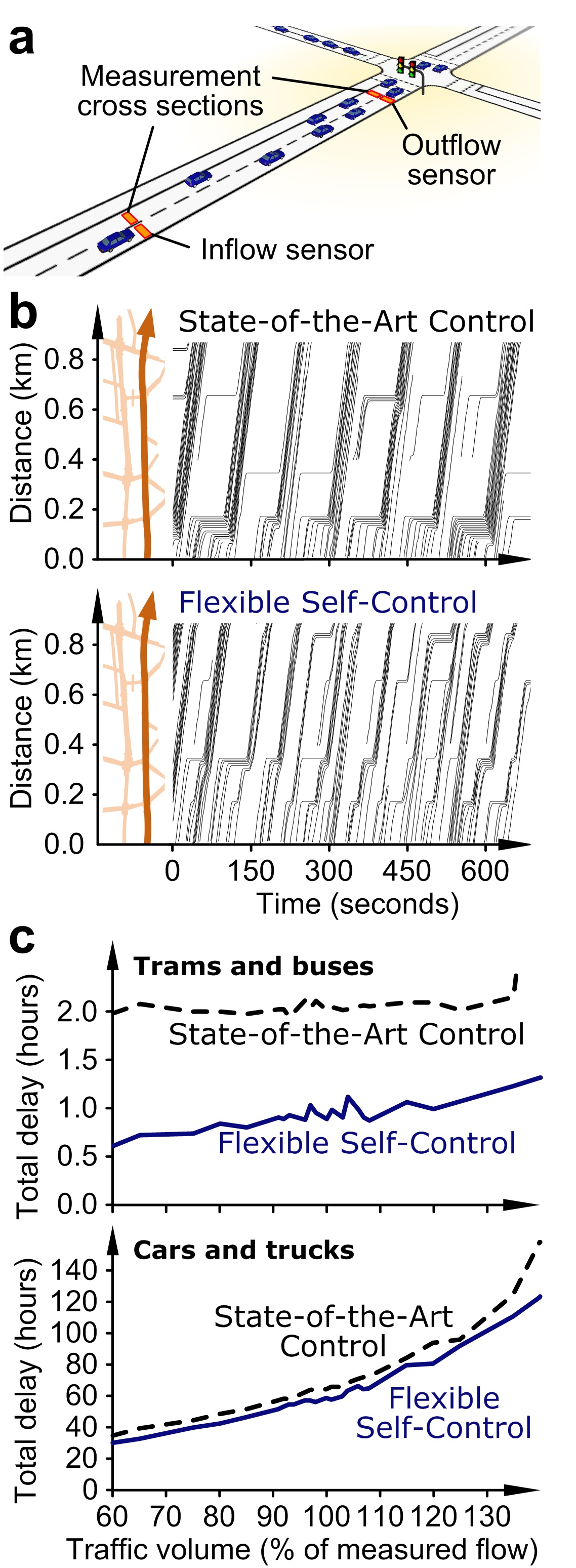,width=2.5in}} \caption{(a)
Traffic intersection with an inflow sensor to anticipate the arrival
flow and an outflow sensor to determine the time when queues have
resolved. Alternatively, one could use recently developed sensors,
which are measuring traffic flows from above. (b) Vehicle
trajectories along an arterial road. Top: State-of-the-art traffic
control, implementing green waves based on a cyclical traffic light
control. Bottom: Identical situation, but traffic lights are
operated according to a self-organized control principle (see also video
SV4). The resulting vehicle platoons are shorter and more irregular,
so that traffic lights can use occurring gaps in the competing or
intersecting flows and flexibly adjust to variations in the flows.
(c) Resulting reduction in the total delay of trams or buses (top)
and of other vehicles (bottom) as a function of the traffic volume.
\label{fig3}}
\end{figure}
Coordinating vehicle flow and traffic lights
in cities is a typical example for this problem
\cite{r16_laemmer16} (see Fig. \ref{fig3}): At each traffic light, the order, duration,
and starting times of green phases can be varied, resulting in a
combinatorially large variety of possible solutions. The
coordination of traffic streams is usually reached by restricting to
cyclical signal operation, and by synchronizing these cycles.
Optimal control strategies for an urban area are normally determined
off-line, based on typical traffic conditions (say, for morning or
evening rush hours). The control centre switches between different
standard control strategies, depending on the traffic conditions.
Complementary, green time durations may be adjusted locally.
However, the traffic conditions, for which the signal schemes were
optimized, never occur exactly. In fact, large fluctuations in the
number of vehicles arriving during one cycle time make the traffic
situation hard to predict. This can affect traffic performance
considerably, in particular if road networks are heterogeneous and
public transport is prioritized, which is normally the case. The
problem is, therefore, to optimize a dynamic and strongly varying
system, which can only be predicted over short time periods and
reacts to control attempts very sensitively. Moreover, operating the
system at maximum performance (capacity), as an optimization of
throughput suggests, makes it vulnerable to breakdowns
\cite{r2_helbing2}: Delayed responses and dynamic feedbacks can
trigger large-scale, cascade-like congestion spreading, i.e.
systemic impacts, as perturbations in the flow of arriving vehicles
may cause jammed road sections, which reduces the intersection
capacity upstream and eventually produces gridlocks. 

Problems like
this occur in many complex systems, and the approach of BioLogistics
can complement established Operations Research methods by revealing
new heuristics for such challenges. One successful method suited for
traffic light control, for example, is based on a pressure principle
that produces self-organized oscillations. This method works as
follows: Based on measurements of the inflows into the road sections
(see Fig.~\ref{fig3}a), one anticipates the vehicle flows that will
arrive at the traffic lights a short time later. These anticipated
flows allow one to determine the expected delays to all traffic
streams, imposed by the respective traffic lights. Then, ``traffic
pressures'' are defined by the temporal increase of the
stream-specific cumulative delays, and at each intersection a green
light is given to the traffic stream that exerts the highest traffic
pressure. According to this principle, traffic streams control the
traffic lights rather than the other way round. To avoid
instabilities, growing queues are stabilized, whenever needed. 

The resulting control principle is self-organized and decentralized
(bottom-up approach). It reacts flexibly to the actual local
situation rather than an average situation, and the short-term
anticipation of vehicle flows reaches a coordination of neighbouring
traffic lights and vehicle streams (see Fig.~\ref{fig3}b). In fact,
self-control does not fight fluctuations in the flow by imposing a
certain flow rhythm. It uses randomly appearing gaps in the flow to
serve other traffic streams. According to simulation studies, this
principle can reduce average delay times by 10 to 30\% (see
Fig.~\ref{fig3}c). The variation in travel times goes down as well,
although the signal operation tends to be non-periodic and,
therefore, less predictable. Maximum acceptable red times and
super-critical queue lengths are taken into account. Pedestrians are
considered by virtual arrival flows, and trams or buses get a
greater weight than vehicles.
\end{itemize}
}
In complex systems, it is often advantageous to support
principles of {\em self-control} and {\em self-stabilization}, as is
done by forthcoming traffic assistant systems \cite{r17_kesting17}.
Given the short-term predictability of the systems we are most
interested in, strong, centralized control would potentially waste
an unnecessarily high amount of resources for information gathering,
transmission, processing and control. It may also affect the
adaptation of the system to changed conditions: If control is too
forceful, the self-organization of autonomously operating elements,
which naturally results from non-linear interactions, is overruled
rather than efficiently used \cite{r14_mikhailov14}. In order to
profit from self-organization, a certain degree of variability must
be tolerated to allow for flexible operation. In fact, rather than
steadily forcing the system not to deviate from its planned
behaviour, which is costly, biology seems to apply the principle of
{\em ``guided self-organization''} or {\em ``moderated
self-control''} \cite{r18_helbing18}: If the interactions between
the system elements are suitable, only small feedback signals are
necessary to reach the desired behaviour. Hence, rather than strict
forcing by a higher hierarchical level, the success principle is
gentle interference (see, for example, the principle of ``chaos
control'') \cite{r13_kaneko13,r14_mikhailov14}. This is usually
based on the combination of antagonistic mechanisms, such as
autocatalytic and inhibitory processes, or the combination of birth
and death processes like fission and fusion (see Fig.~\ref{fig2}).
Antagonistic mechanisms are particularly important for growth and
size control. If they are not well balanced, the system can collapse
or fail (see video SV3), or parts of it can expand in an
uncontrolled way (cf. the example of cancer). The grand challenge of
creating systems based on ``guided self-organization'' is the {\em
design of suitable interaction mechanisms} between the system
elements. For example, given a desired final structure, how should
the parts and interactions in a self-assembly system be chosen?
Self-organization per se does not cause optimal outcomes.

It must be avoided that the system gets stuck in a suboptimal,
``frustrated'' state, or that it behaves unstable, which could cause
local breakdowns or even systemic failures. Some experience has
recently been gained through the study of self-assembly
\cite{r12_li12}, of self-organization in vehicle traffic and crowds
\cite{r2_helbing2}, or through the design of peer-to-peer (P2P)
systems \cite{r18_helbing18}. Here, various kinds of coordination
can spontaneously emerge by simple interactions
\cite{r19_tumer19,r13_kaneko13,r14_mikhailov14}. Well-known examples
are the synchronization of oscillatory processes or the
establishment of cooperation between agents as studied by game
theory \cite{r18_helbing18}.

\par
Considering all this, focussing on bio-inspired solutions in
logistic systems will establish a fruitful new research field,
addressing challenges of complex systems characterized by non-linear
interactions of their constituents, delays, and possible breakdowns,
where demand and other boundary conditions strongly fluctuate and
are predictable only over short time periods. Circumstances like
these are quite characteristic for economic and logistic systems on
a national and global scale. Based on a better understanding of
logistic design and operation principles in biology, it will be
possible to improve engineered logistics by applying success
principles of nature. BioLogistics also promises concepts for the
construction of bio-logistical systems in bio-nano-engineering
\cite{r20_tatson20,r21_hess21}, as is illustrated by Example 4. It is within
reach to compose nanostructures, using intracellular transport and
production mechanisms in synthetic cell-free environments. Moreover,
it appears realistic to design structures by the cooperation of a
large number of simple machines or robots
\cite{r22_floreano22,r23_krieger23,r14_mikhailov14,r19_tumer19}.

{\small\sffamily\begin{itemize}
\item[] \textbf{Example 4: NanoLogistics}
\par
Molecular shuttles based on the motor proteins and cytoskeletal
filaments have the potential to extend the lab-on-a-chip paradigm to
nanofluidics by enabling the active, directed and selective
transport of molecules and nano-particles
\cite{r33_heuvel33,r34_goel34}. In com-parison with conventional
nano-transport and nanomanipulation devices, bio-molecular motors
(i) are small and can therefore operate in a highly parallel manner,
(ii) are easy to produce and can be modified by genetic engineering,
and (iii) operate with high energy efficiency. One might envision
that biomolecular motors could be used as molecule-sized robots (see
Fig.~\ref{fig4}) that (a) work in molecular factories, where small,
but intricate structures are made on tiny assembly lines, (b)
construct networks of molecular conductors and transistors for use
as electrical circuits, or (c) continually patrol inside
``adaptive'' materials and repair them when necessary. Thus
biomolecular motors could form the basis of bottom-up approaches for
the construction, active structuring and maintenance at the
nanometre scale. 

In addition, artificial nanotransport systems
provide an opportunity to simulate 'real world' traffic and logistic
scenarios. As such, it is conceivable to use molecular transport
systems for the optimization of traffic routing or as biocomputation
tools based on a combinatorial approach \cite{r35_nicolau35}.
\par\begin{figure}[htbp]
\centerline{\psfig{file=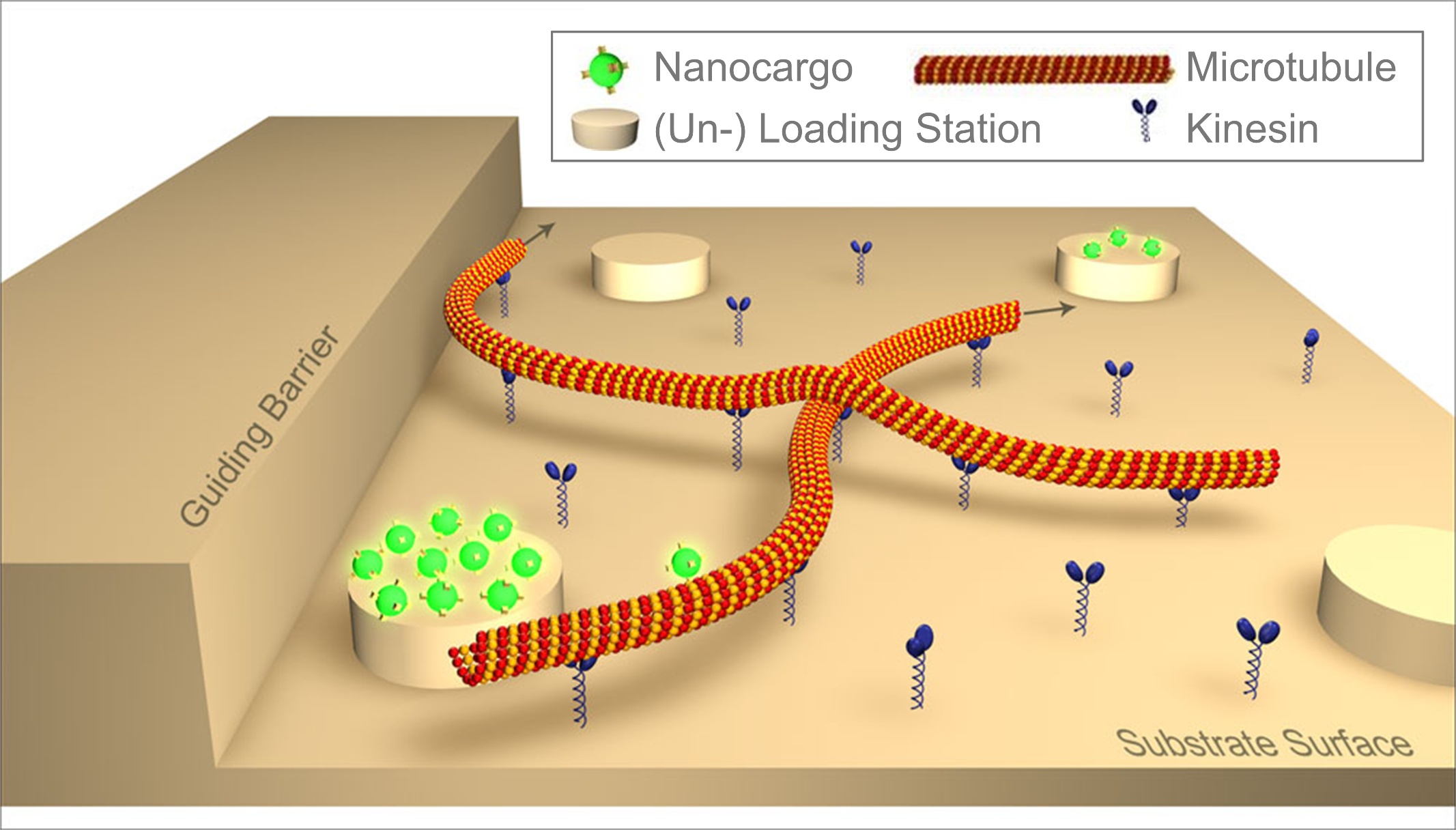,width=4in}}
\caption{Schematic illustration of a possible biomolecular transport
system on the nanoscale. Microtubules are propelled over the surface
of a silicon chip by immobilized kinesin motors (adapted from
Ref.~\cite{r36_kerssemakers36}). Via specific linker molecules, the
modular microtubule transporters may pick up cargo at a loading
station and transport it to another location for unloading or
sorting. To reliably guide the microtubule movement, chemical and
topographical surface structures can be applied, the geometries of which mark the
roads along which the flow of cargo is directed 
(see also video SV5). \label{fig4}}
\end{figure}
\end{itemize}
}

\section{Modular designs}

Biological systems actually use large numbers of simple and small
building blocks that are ``cheap'' to produce
\cite{r14_mikhailov14}. This facilitates modular designs
\cite{r24_hartwell24}, which do not require a {\em specific} unit to
do a certain job: Any equivalent unit can do, which implies
substitutability and robustness. Such modular logistic designs,
which can reorganize quickly, become applicable to real problems
now: Besides large-scale machines and equipment placed in fixed
locations, recently, small multi-functional and mobile elements have
become available to design future logistic systems. Their greater
flexibility results from the combinatorially large number of
functions that can be performed in a cooperative manner.
Furthermore, recent sensor and communication technologies allow
engineers to implement adaptive self-control principles
\cite{r19_tumer19,r25_scholz25,r26_huelsmann26}. One successful
example is the self-organized control of traffic lights
\cite{r16_laemmer16}. Cooperative modular designs addressing future
logistic problems could also put principles of swarm intelligence
\cite{r27_bobabeau27} into practise, as these can create complex and
robust cooperative behaviour from simple interactions. So far,
logistics has not applied these concepts on a large scale. While
solving related logistic challenges of the future, BioLogistics is
expected to make many innovative contributions to these fields --
for example when creating self-controlled systems, in which machines
and robots perform cooperative tasks
\cite{r22_floreano22,r23_krieger23}. In essence, non-linear
interactions between the elements of complex systems pose great
challenges for optimization and control. At the same time, however,
they provide interesting new perspectives for self-stabilization and
self-control.

\section*{Acknowledgments}

The authors are grateful to the Gottlieb Daimler and Karl Benz
Foundation for providing the intellectually stimulating environment
and the financial support for a consortium to work on this project.
They would like to thank Alexander Mikhailov for stimulating
discussions on related subjects. Cecile Leduc, Vladimir Varga and
Jonathon Howard are acknowledged for discussions and experimental
contributions in the field of motor interactions with microtubules.
Furthermore, partial financial support was provided by the German
Research Foundation (DFG projects He 2789/5-1, 8-1) and the
Volkswagen Foundation (project I/82 697) as well as the Max Planck
Society, the German Ministry for Education and Research (systems
biology network HepatoSys grant 0313082J, and grant 03 N 8712), and
the ETH Competence Center 'Coping with Crises in Complex
Socio-Economic Systems' (CCSS) through ETH Research Grant
CH1-01-08-2.

\section*{Author Information}
Correspondence and requests for materials
should be addressed to D.H. (dhelbing@ethz.ch) and M.Z.
(zerial@mpg-cbg.de).

\section*{Supplementary Information}
The supplementary videos SV1 to SV5 and the supplementary table 1 are available at 
\verb|http://www.soms.ethz.ch/research/biologistics|
\clearpage



\begin{thebibliography}{99}

\bibitem{r1_anderson1}
Anderson, P. W., More is different, {\em Science} {\bf 177} (1972)
393--396.
\bibitem{r31_aridor31} Aridor, M. and Hannan, L. A., Traffic jam: A compendium of human diseases that affect intracellular transport processes, {\em Traffic} {\bf 1} (2000) 836--851.
\bibitem{r3_ball3} Ball, P., Life's lessons in design, {\em Nature} {\bf 409} (2001) 413--416.
\bibitem{r9_banavar9} Banavar, J. R., Maritan, A. and Rinaldo, A., Size and form in efficient transportation networks, {\em Nature} {\bf 399} (1999) 130--132.
\bibitem{r27_bobabeau27} Bonabeau, E., Dorigo, M. and Theraulaz, G., Inspiration for optimization from social insect behaviour, {\em Nature} {\bf 406} (2000) 39--42.
\bibitem{r4_dickinson4} Dickinson, M. H., Bionics: Biological insight into mechanical design, {\em Proc. Natl. Acad. Sci. USA} {\bf 96} (1999) 14208--14209.
\bibitem{r30_dixit30} Dixit, R., Ross, J. L., Goldman, Y. E. and Holzbaur, E. L. F., Differential regulation of dynein and kinesin motor proteins by tau, {\em Science} {\bf 319} (2008) 1086--1089.
\bibitem{r22_floreano22} Floreano, D. and Mattiussi, C., {\em Bio-Inspired Artificial Intelligence: Theories, Methods, and Technologies} (MIT Press, Cambridge, 2008).
\bibitem{r34_goel34} Goel, A. and Vogel, V., Harnessing biological motors to engineer systems for nanoscale transport and assembly, {\em Nature Nanotechnol.} {\bf 3} (2008) 465--475.
\bibitem{r24_hartwell24} Hartwell, L. H., Hopfield, J. J., Leibler, S. and Murray, A. W., From molecular to modular cell biology, {\em Nature} {\bf 402} Supp. (1999) C47--C52.
\bibitem{r2_helbing2} Helbing, D., Traffic and related self-driven many-particle systems, {\em Rev. Mod. Phys.} {\bf 73} (2001) 1067--1141.
\bibitem{r18_helbing18} Helbing, D. (ed.) {\em Managing Complexity} (Springer, Berlin, 2008).
\bibitem{r10_helenius10} Helenius, J., Brouhard, G., Kalaidzidis, Y., Diez, S. and Howard, J., The depolymerizing kinesin MCAK uses lattice diffusion to rapidly target microtubule ends, {\em Nature} {\bf 441} (2006) 115--119.
\bibitem{r21_hess21} Hess, H., Toward devices powered by biomolecular motors, {\em Science} {\bf 312} (2006) 860--861.
\bibitem{r26_huelsmann26} H{\"u}lsmann, M. and Windt, K. (eds.) {\em Understanding Autonomous Cooperation and Control in Logistics} (Springer, Berlin, 2007).
\bibitem{r13_kaneko13} Kaneko, K., {\em Life: An Introduction to Complex Systems Biology} (Springer, Berlin, 2006).
\bibitem{r36_kerssemakers36} Kerssemakers, J. et al., 3D nanometer tracking of motile microtubules on reflective surfaces, {\em Small} {\bf 5}(15) (2009) 1732--1737.
\bibitem{r17_kesting17} Kesting, A., Treiber, M., Sch{\"o}nhof, M. and Helbing, D., Adaptive cruise control design for active congestion avoidance, {\em Transportation Research C} {\bf 16}(6) (2008) 668--683.
\bibitem{r28_korten28} Korten, T. and Diez, S., Setting up roadblocks for kinesin-1: mechanism for the selective speed control of cargo carrying microtubules, {\em Lab Chip} {\bf 8} (2008) 1441--1447.
\bibitem{r23_krieger23} Krieger, M. J. B., Billeter, J.-B. and Keller, L., Ant-like task allocation and recruitment in cooperative robots, {\em Nature} {\bf 406} (2000) 992--995.
\bibitem{r6_kumar6} Kumar, S. and Bentley, P. (eds.) {\em On Growth, Form and Computers} (Academic Press, London, 2003).
\bibitem{r16_laemmer16} L{\"a}mmer, S. and Helbing, D., Self-control of traffic lights and vehicle flows in urban road networks, {\em J. Stat. Mech.} (2008) P04019.
\bibitem{r12_li12} Li, M., Schnablegger, H. and Mann, S., Coupled synthesis and self-assembly of nanoparticles to give structures with controlled organization, {\em Nature} {\bf 402} (1999) 393--395.
\bibitem{r14_mikhailov14} Mikhailov, A. S. and Calenbuhr, V., {\em From Cells to Societies} (Springer, Berlin, 2006).
\bibitem{r35_nicolau35} Nicolau, D. V. et al., Molecular motors-based micro- and nano-biocomputation devices, {\em Microelectronic Engineering} {\bf 83} (2006) 1582--1588.
\bibitem{r7_passino7} Passino, K. M., {\em Biomimicry for Optimization, Control, and Automation} (Springer, London, 2005).
\bibitem{r32_rink32} Rink, J., Ghigo, E., Kalaidzidis, Y. and Zerial, M., Rab conversion as a mechanism of progression from early to late endosomes, {\em Cell} {\bf 122} (2005) 735--749.
\bibitem{r25_scholz25} Scholz-Reiter, B., Windt, K. and Freitag, M., Autonomous logistic processes -- New demands and first approaches, {\em Proc. 37th CIRP-ISMS} (2004) 357--362.
\bibitem{r20_tatson20} Taton, T. A., Two-way traffic, {\em Nature Mater.} {\bf 2} (2003) 73--74.
\bibitem{r19_tumer19} Tumer, K. and Wolpert, D. (eds.) {\em Collectives and the Design of Complex Systems} (Springer, New York, 2004).
\bibitem{r15_ueda15} Ueda, K., Vaario, J. and Ohkura, K., Modelling of biological manufacturing systems for dynamic reconfiguration, {\em Annals of the CIRP} {\bf 46} (1997) 343--346.
\bibitem{r33_heuvel33} van den Heuvel, M. G. and Dekker, C., Motor proteins at work for nanotechnology, {\em Science} {\bf 317} (2007) 333--336.
\bibitem{r29_varga29} Varga, V. et al., Yeast kinesin-8 depolymerizes microtubules in a length-dependent manner, {\em Nature Cell Biol.} {\bf 8} (2006) 957--962.
\bibitem{r5_wadhawan5} Wadhawan, V. K., {\em Smart Structures} (Oxford University Press, Oxford, 2007).
\bibitem{r8_west8} West, G. B., Brown, J. H. and Enquist, B. J., The fourth dimension of life: Fractal geometry and allometric scaling of organisms, {\em Science} {\bf 284} (1999) 1677--1679.
\bibitem{r11_zhang11} Zhang, S., Fabrication of novel biomaterials through molecular self-assembly, {\em Nature Biotechnol.} {\bf 21} (2003) 1171--1178.

\end{thebibliography}
\end{document}